\documentclass[10pt]{iopart}

\expandafter\let\csname equation*\endcsname\relax
\expandafter\let\csname endequation*\endcsname\relax
\usepackage{amsmath,amssymb,amsfonts}
\usepackage{graphicx}
\usepackage{bm}
\usepackage{hyperref}
\usepackage{xcolor}
\usepackage{braket}
\usepackage{booktabs}
\usepackage{tikz}
\usetikzlibrary{arrows.meta, decorations.pathreplacing, calc, positioning}

\begin{document}

\title{Directional Dynamics of the Non-Hermitian Skin Effect}

\author{Bin Yi$^1$}
\address{$^1$ Institute of Fundamental and Frontier Sciences, University of Electronic Science and Technology of China, Chengdu 610051, China}
\ead{ucapbyi@uestc.edu.cn}


\begin{abstract}
The dynamical consequences of the non-Hermitian skin effect (NHSE) remain largely unexplored despite extensive studies of its static properties. Here we address this gap by applying quantum Liang information flow (QLIF)---an inherently directional measure of causal influence---to the non-Hermitian Su-Schrieffer-Heeger model with non-reciprocal hopping. Unlike symmetric correlation functions, QLIF directly captures the directional asymmetry $\mathbb{T}_{R \to L} \neq \mathbb{T}_{L \to R}$ characteristic of non-reciprocal systems. We demonstrate a ``scissors effect'' where the asymmetry $\Delta_{\mathbb{T}}$ varies approximately linearly with the non-reciprocity parameter $\gamma$ for small $|\gamma|$, and exhibits non-monotonic dependence on the skin length $\xi$, with optimal asymmetry at moderate skin localization. The velocity ordering $v_{\mathrm{eff}}(\gamma < 0) > v_{\mathrm{eff}}(0) > v_{\mathrm{eff}}(\gamma > 0)$ reveals NHSE-induced blocking of information flow against the skin direction. Three distinct temporal regimes emerge: light-cone-bounded spreading, $\gamma$-dependent stabilization, and coherent oscillations. These results establish the first quantitative connection between static skin localization and directional information dynamics, offering new insights into information propagation in non-reciprocal quantum systems.
\end{abstract}

\pacs{03.65.Vf, 11.30.Er, 05.60.Gg, 73.20.Fz}

\maketitle


\section{Introduction}
\label{sec:intro}

The non-Hermitian skin effect (NHSE)---the dramatic accumulation of all eigenstates at system boundaries under open boundary conditions---has emerged as a defining phenomenon in non-Hermitian physics~\cite{Bergholtz2021,Okuma2023,Gong2018,Ashida2020,Kawabata2019,YaoWang2018,HatanoNelson1996,Kunst2018,Lee2016,Okuma2020,Zhang2020}. Experimental realizations now span photonic lattices~\cite{Weidemann2020,Wang2021Science}, topolectrical circuits~\cite{Helbig2020,Hofmann2020}, quantum walks~\cite{LiZhangYi2021CPL,Xiao2020}, and ultracold atoms~\cite{Zhao2025Nature}. While the relationship between non-reciprocal couplings and skin accumulation direction is well established in one-dimensional models~\cite{YaoWang2018,Okuma2020,Zhang2020}, recent works have shown that this correspondence can be nontrivially modified or even reversed in more general settings~\cite{YangLee2025,Roccati2024,ShiPoddubny2025}. However, existing studies have primarily focused on \emph{static} properties---spectra, topological invariants, and eigenstate distributions~\cite{WangGuoDuKou2020CPL,Lin2023,Okuma2023,Lee2019,Herviou2019,Longhi2019,Yokomizo2019}---while the \emph{dynamical} consequences remain comparatively unexplored~\cite{WangLi2025CPL,Song2019,Xue2022,Xiao2024,Li2024Dynamic,Yoshida2024Mott,Shimomura2024Fock,Zhang2025Algebraic,Zeng2022,Li2020}. A central open question is how the non-reciprocity parameter $\gamma$ affects information propagation.

Traditional dynamical probes are inadequate for this task. Time-dependent correlation functions $\langle \sigma_i(t) \sigma_j(0) \rangle$ are inherently symmetric and cannot capture directional bias~\cite{HatanoNelson1996,Song2019}. The Lieb-Robinson bound~\cite{LiebRobinson1972,Bravyi2006}, which constrains information spreading in Hermitian systems, breaks down in the non-Hermitian case~\cite{Barch2024}---as demonstrated by supersonic modes observed on trapped-ion quantum computers~\cite{Zhang2025Supersonic}. The quasiparticle picture~\cite{Calabrese2005} also faces challenges with complex spectra and biorthogonal eigenstates~\cite{Kunst2018,Borgnia2020}. What is needed is a measure that directly quantifies \emph{directional asymmetry} in information flow.

Quantum Liang information flow (QLIF)~\cite{YiBose2022,Liang2008,Liang2016} provides precisely such a measure: $T_{B \to A} = dS_A/dt - dS_{A\!\not\! B}/dt$, where $S_{A\!\not\! B}$ denotes entropy evolution with subsystem $B$ frozen~\cite{ChenZhouChenYe2024CPL}. Unlike correlations, QLIF is inherently directional ($T_{B \to A} \neq T_{A \to B}$), making it ideally suited for non-reciprocal systems~\cite{Ghosh2025}. Here we apply QLIF to the non-Hermitian SSH model~\cite{SSH1979} with non-reciprocal hopping $t_1 \pm \gamma$. We discover: (i) a ``scissors effect'' where $\Delta_{\mathbb{T}} = \mathbb{T}_{R \to L} - \mathbb{T}_{L \to R}$ varies linearly with $\gamma$ for small $\gamma$; (ii) non-monotonic dependence on skin length $\xi$; (iii) velocity ordering $v_{\mathrm{eff}}(\gamma < 0) > v_{\mathrm{eff}}(0) > v_{\mathrm{eff}}(\gamma > 0)$ revealing NHSE-induced blocking; and (iv) three distinct temporal regimes. These results establish the first quantitative link between skin localization and directional information dynamics.


\section{Model}
\label{sec:model}

We consider a non-Hermitian Su-Schrieffer-Heeger (SSH) chain with $N$ unit cells under open boundary conditions~\cite{YaoWang2018}:
\begin{equation}
    H = \sum_{j=1}^{N} \mathbf{c}_j^\dagger \mathcal{T}_0 \, \mathbf{c}_j
    + \sum_{j=1}^{N-1} \left( \mathbf{c}_{j+1}^\dagger \mathcal{T}_+ \, \mathbf{c}_j + \text{H.c.} \right),
    \label{eq:H_matrix}
\end{equation}
where $\mathbf{c}_j = (c_{\alpha,j}, c_{\beta,j})^T$ and
\begin{equation}
    \mathcal{T}_0 = \begin{pmatrix} 0 & t_1 + \gamma \\ t_1 - \gamma & 0 \end{pmatrix}, \quad
    \mathcal{T}_+ = \begin{pmatrix} 0 & 0 \\ t_2 & 0 \end{pmatrix}.
    \label{eq:hopping_matrices}
\end{equation}
Here $t_1$ ($t_2$) is the intracell (intercell) hopping, and $\gamma$ parametrizes non-reciprocity~\cite{Bergholtz2021}.

For $\gamma \neq 0$, the non-Hermitian skin effect (NHSE) causes all eigenstates to localize exponentially at one boundary~\cite{YaoWang2018,Kunst2018}. The skin parameter $r = \sqrt{|(t_1 - \gamma)/(t_1 + \gamma)|}$ determines localization: $\gamma > 0$ ($r < 1$) gives left-boundary localization; $\gamma < 0$ ($r > 1$) gives right-boundary localization. The skin length $\xi = 1/|\ln r|$ diverges in the Hermitian limit.

We set $t_1 = 1$, $t_2 = 0.5$, and $L = 42$ sites, scanning $\gamma$ across the full physically allowed range $|\gamma| < t_1$. Details on the bulk spectrum and light-cone velocity are in the Supplemental Material~\cite{SM}.


\section{Quantum Liang Information Flow}
\label{sec:qlif}

\begin{figure}[htbp]
    \centering
    \includegraphics[width=0.85\columnwidth]{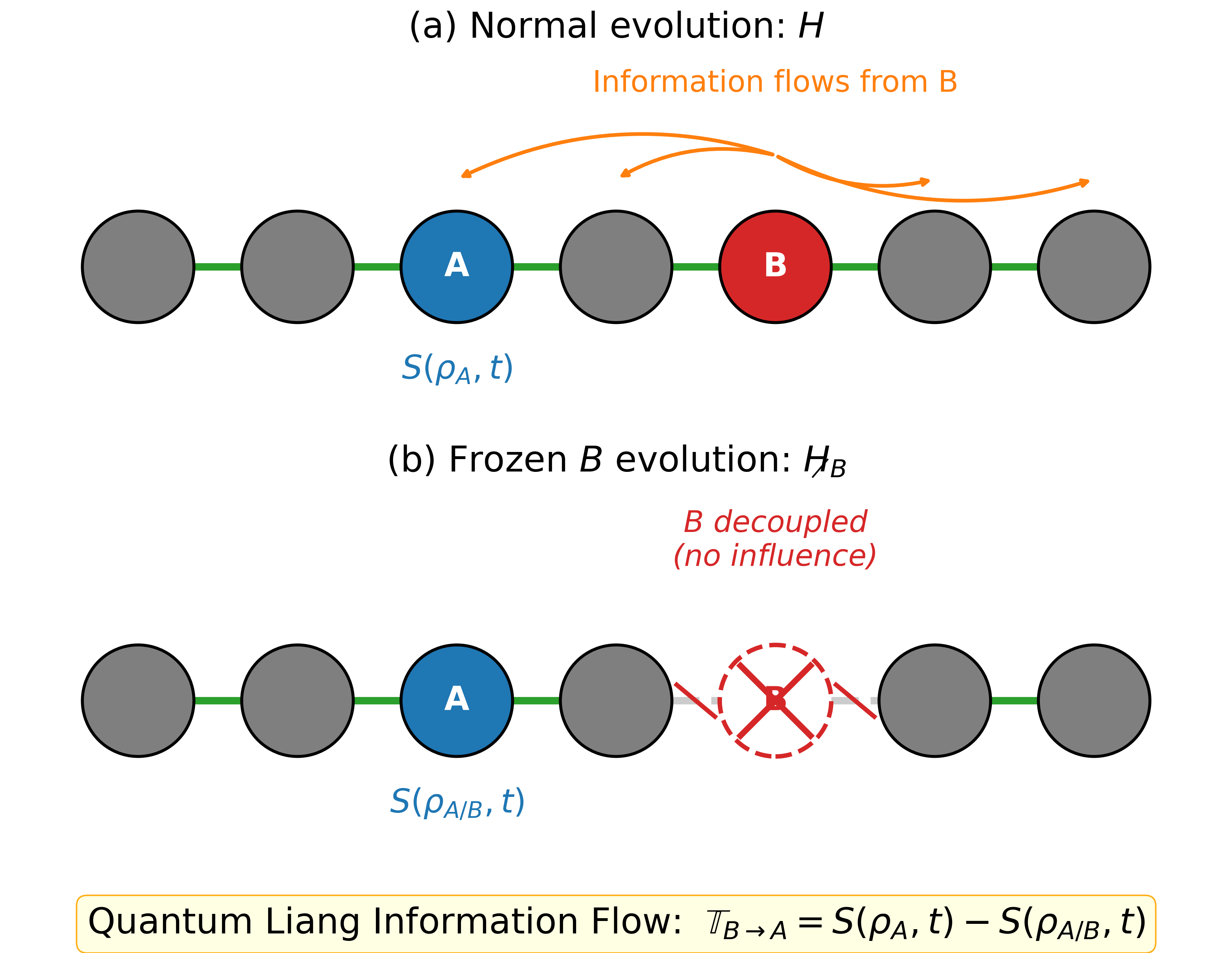}
    \caption{Freezing operation for QLIF. (a) Normal evolution: all couplings active. (b) Frozen-$B$ evolution: couplings to $B$ removed. The QLIF is the entropy difference between (a) and (b).}
    \label{fig:freezing}
\end{figure}

The quantum Liang information flow (QLIF)~\cite{YiBose2022,Liang2008,Liang2016,Ghosh2025} quantifies the causal influence of subsystem $B$ on $A$ through a freezing operation. Let $S_A(t) = -\tr[\rho_A \ln \rho_A]$ be the von Neumann entropy of $A$. We construct a modified Hamiltonian $H_{\!\not\! B}$ with all couplings to $B$ removed, and define the cumulative QLIF as
\begin{equation}
    \mathbb{T}_{B \to A}(t) = S_A(t) - S_{A\!\not\! B}(t),
    \label{eq:cumulative_qlif}
\end{equation}
where $S_{A\!\not\! B}$ is the entropy under frozen-$B$ evolution (Fig.~\ref{fig:freezing}). Unlike correlation functions, QLIF is inherently directional: $\mathbb{T}_{B \to A} \neq \mathbb{T}_{A \to B}$ in general.

We place the initial state at chain center $j_0$, with observation sites $A$ (left) and $B$ (right) at equal distance $d$. The directional asymmetry $\Delta_{\mathbb{T}} = \mathbb{T}_{R \to L} - \mathbb{T}_{L \to R}$ vanishes for $\gamma = 0$ and becomes nonzero when non-reciprocity breaks inversion symmetry. Implementation details are in the Supplemental Material~\cite{SM}.


\section{Results}
\label{sec:results}

We compute the cumulative QLIF for a single particle initially localized at the chain center, with observation sites at equal distance $d$ on opposite sides. All sites reside on the same sublattice to preserve inversion symmetry at $\gamma = 0$ (see Supplemental Material~\cite{SM}).

\begin{figure}[htbp]
    \centering
    \includegraphics[width=0.75\columnwidth]{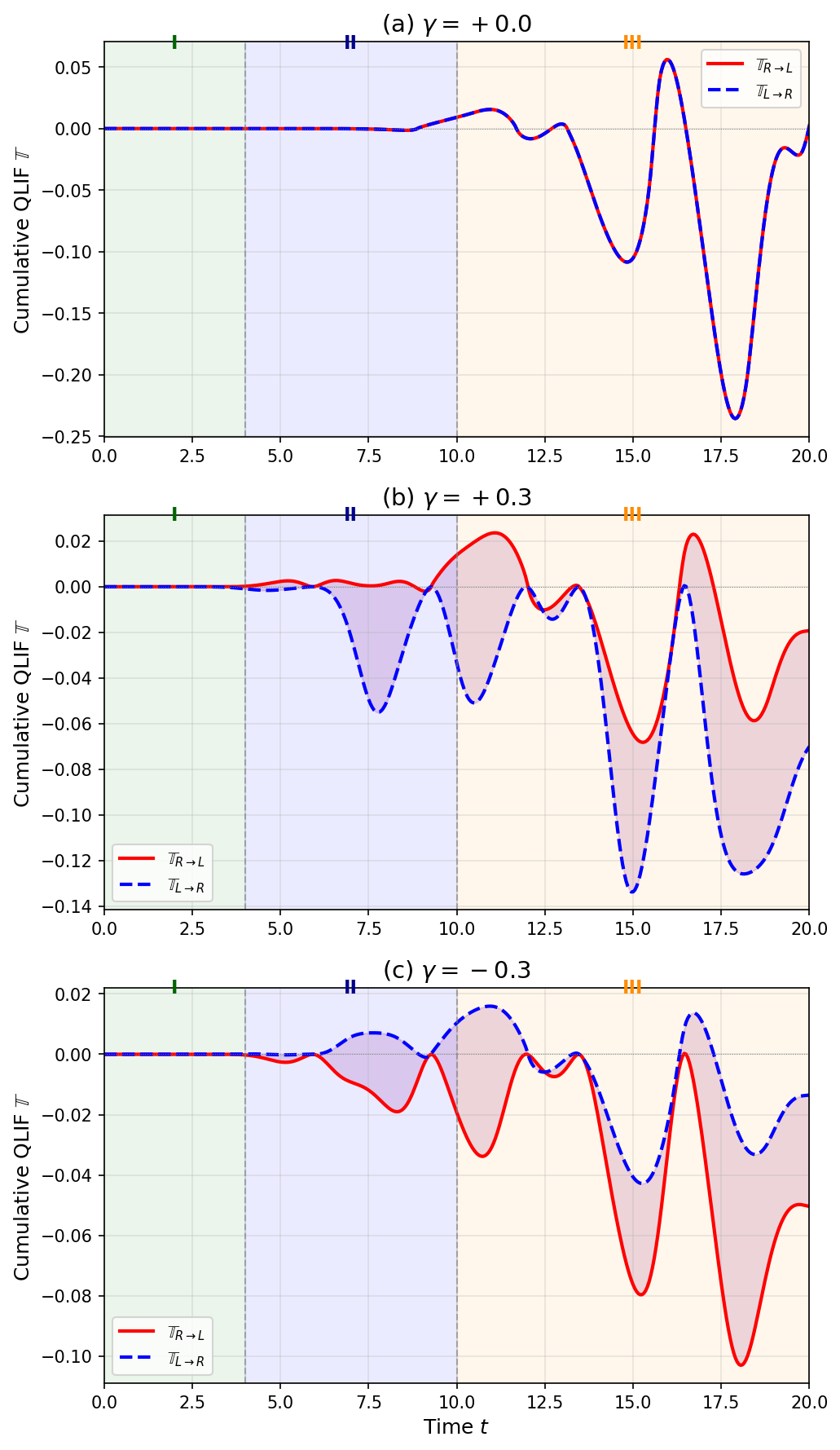}
    \caption{Scissors effect in cumulative QLIF. Time evolution of $\mathbb{T}_{R \to L}$ (solid) and $\mathbb{T}_{L \to R}$ (dashed) for (a) $\gamma = 0$ (Hermitian, perfect overlap), (b) $\gamma = +0.3$, and (c) $\gamma = -0.3$. The shaded region highlights the directional asymmetry. Three temporal regimes are visible: (I) onset, (II) stabilization, (III) oscillations. Parameters: $L = 42$, $d = 6$, $t_1 = 1$, $t_2 = 0.5$.}
    \label{fig:scissors}
\end{figure}

\paragraph{Scissors effect.}
\label{sec:scissors}
Our first key result concerns the breakdown of left-right symmetry in QLIF when non-reciprocity is introduced. In the Hermitian limit ($\gamma = 0$), spatial inversion symmetry ensures $\mathbb{T}_{R \to L}(t) = \mathbb{T}_{L \to R}(t)$. For $\gamma \neq 0$, this symmetry is broken. Figure~\ref{fig:scissors} displays the ``scissors'' pattern---where the two curves diverge after an initial overlap---for $\gamma = \pm 0.3$. Panel (a) confirms perfect overlap in the Hermitian case, with deviations at the level of machine precision ($< 10^{-14}$). Panels (b) and (c) reveal the symmetry breaking: the directional asymmetry $\Delta_{\mathbb{T}} = \mathbb{T}_{R \to L} - \mathbb{T}_{L \to R}$ is positive for $\gamma > 0$ (enhanced right-to-left flow, left-boundary skin localization) and negative for $\gamma < 0$. At intermediate times ($t \approx 10$), the cumulative QLIF can be either positive (source increases target entropy) or negative (source decreases target entropy), depending on the interplay between wavepacket spreading and interference effects. This reflects the non-reciprocal intracell hopping: $\gamma > 0$ enhances $\beta \to \alpha$ hopping while suppressing $\alpha \to \beta$, manfiesting macroscopically as directional bias. The sign reversal of $\mathbb{T}$ for opposite $\gamma$ reflects the fundamental directional asymmetry induced by the NHSE.

\begin{figure}[htbp]
    \centering
    \includegraphics[width=0.95\columnwidth]{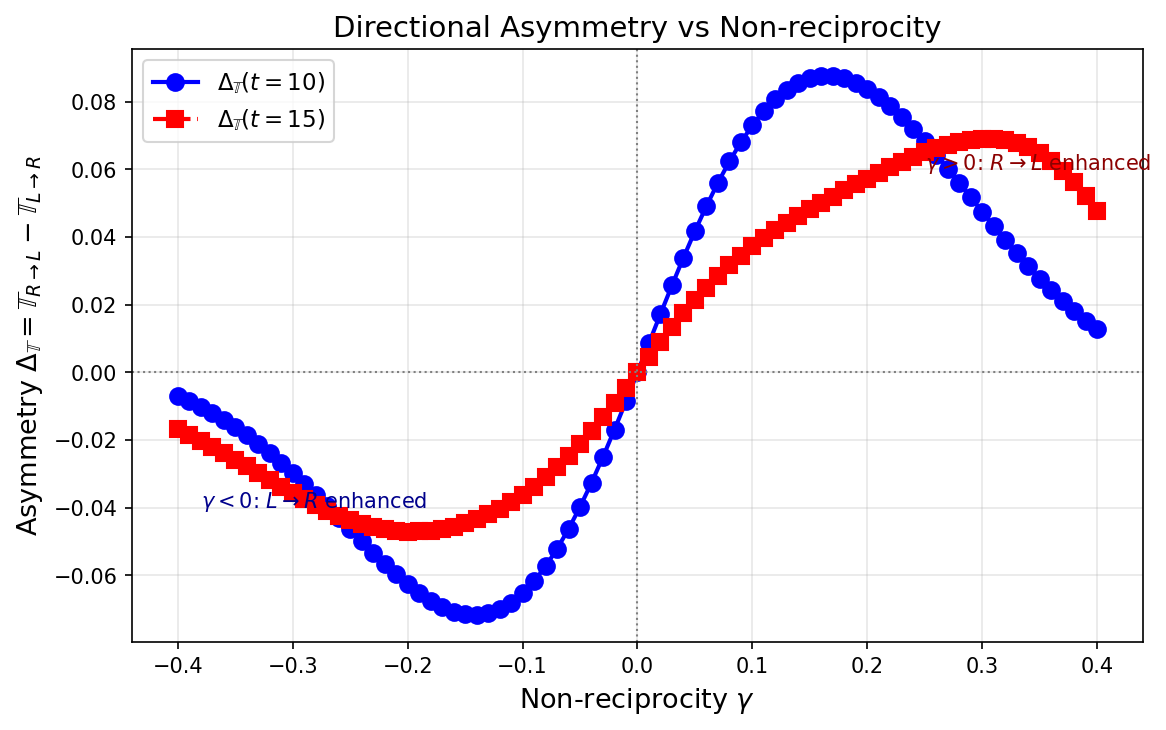}
    \caption{Directional asymmetry $\Delta_{\mathbb{T}}$ versus $\gamma$ at $t = 10$ (circles) and $t = 15$ (squares) across the full range $|\gamma| < t_1$. The sign rule $\mathrm{sgn}(\Delta_{\mathbb{T}}) = \mathrm{sgn}(\gamma)$ holds throughout. The non-monotonic behavior---peaking at moderate $|\gamma|$ and vanishing as $|\gamma| \to t_1$---reflects competition between symmetry breaking and extreme skin localization that confines dynamics to boundaries.}
    \label{fig:delta_gamma}
\end{figure}

To systematically characterize this dependence, we scan across the full physically allowed range $|\gamma| < t_1$ and extract $\Delta_{\mathbb{T}}$ at fixed times. Figure~\ref{fig:delta_gamma} reveals a key result: the sign rule
\begin{equation}
    \mathrm{sgn}(\Delta_{\mathbb{T}}) = \mathrm{sgn}(\gamma)
    \label{eq:sign_rule_delta}
\end{equation}
holds throughout the entire range, directly addressing the open question of ``how the behavior of NHSE is influenced by the variation of nonreciprocity''~\cite{Zeng2022}. Positive $\gamma$ (left-boundary skin localization) enhances right-to-left information flow ($\Delta_{\mathbb{T}} > 0$), while negative $\gamma$ (right-boundary localization) enhances left-to-right flow ($\Delta_{\mathbb{T}} < 0$).

Crucially, the asymmetry exhibits \textit{non-monotonic} $\gamma$-dependence: $|\Delta_{\mathbb{T}}|$ increases approximately linearly for small $|\gamma|$, peaks at moderate values ($|\gamma| \approx 0.15$--$0.3$), then decreases toward zero as $|\gamma| \to t_1$. The suppression at large $|\gamma|$ has a clear microscopic origin: the intracell hopping amplitudes are $t_1 + \gamma$ ($\alpha \to \beta$) and $t_1 - \gamma$ ($\beta \to \alpha$). As $\gamma \to t_1$, one direction becomes completely suppressed ($t_1 - \gamma \to 0$) while the other doubles ($t_1 + \gamma \to 2t_1$), making hopping perfectly unidirectional. The skin length $\xi = 1/|\ln r| \to 0$ in this limit, confining all eigenstates to within one or two boundary sites. Since the initial excitation and observation sites reside in the bulk (distance $d = 6$ from chain center), the signal cannot propagate when $\xi \ll d$. Paradoxically, the strongest NHSE produces the weakest measurable directional asymmetry---a key prediction for experiments seeking optimal operating regimes.

\begin{figure}[htbp]
    \centering
    \includegraphics[width=0.95\columnwidth]{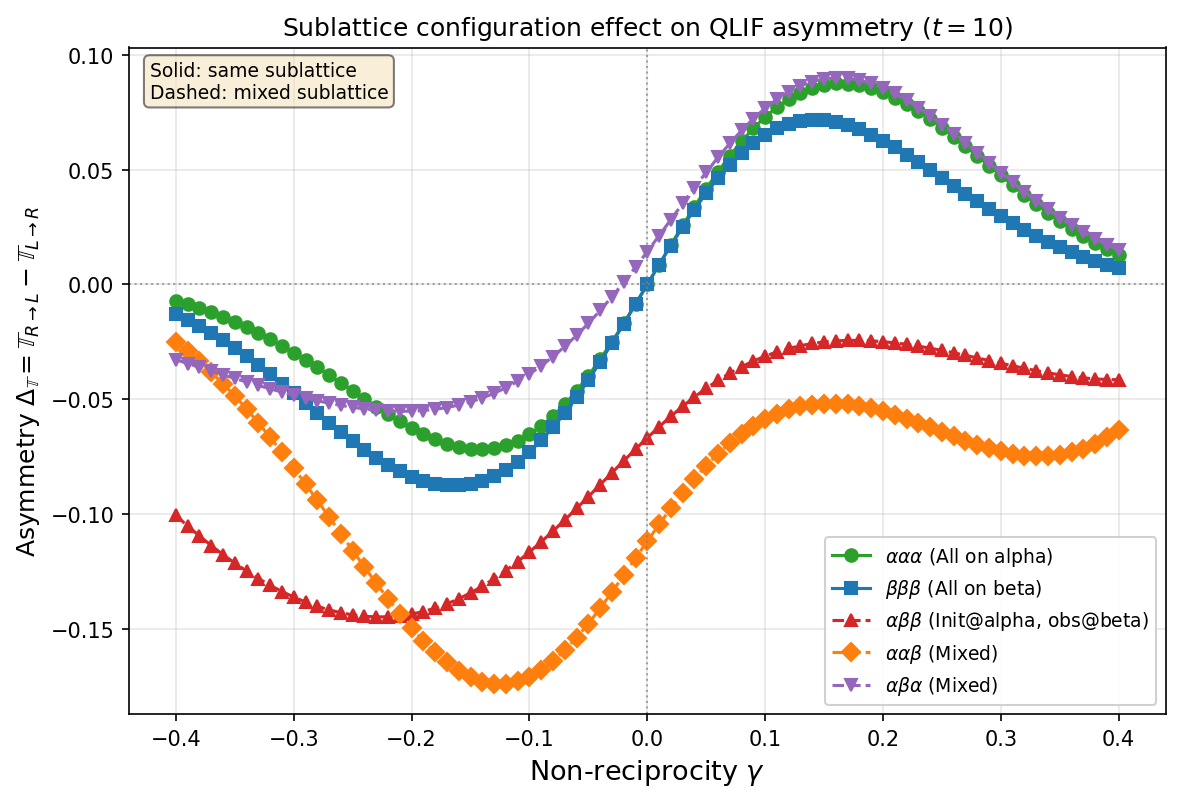}
    \caption{Sublattice configuration effect on QLIF asymmetry. Solid lines: same-sublattice ($\alpha\alpha\alpha$, $\beta\beta\beta$) pass through the origin. Dashed lines: mixed-sublattice configurations show nonzero offset at $\gamma = 0$ due to structural asymmetry from bipartite hopping.}
    \label{fig:sublattice}
\end{figure}

\paragraph{Sublattice configuration.}
\label{sec:sublattice}
The SSH model's bipartite structure imposes crucial constraints on QLIF measurements. Figure~\ref{fig:sublattice} compares five configurations labeled by sublattices of initial state and observation sites. Same-sublattice configurations ($\alpha\alpha\alpha$, $\beta\beta\beta$; solid lines) pass through the origin $(\gamma, \Delta_{\mathbb{T}}) = (0, 0)$ with machine precision, as required by spatial inversion symmetry. Remarkably, the curves satisfy $\Delta_{\mathbb{T}}^{\beta\beta\beta}(\gamma) = \Delta_{\mathbb{T}}^{\alpha\alpha\alpha}(-\gamma)$, reflecting the sublattice exchange symmetry $\alpha \leftrightarrow \beta$ combined with $\gamma \to -\gamma$. Both curves exhibit a pronounced nonlinear response: the asymmetry peaks at moderate $|\gamma|$ before decreasing at larger $|\gamma|$ due to strong localization effects. Mixed-sublattice configrations (dashed lines) exhibit nonzero asymmetry even at $\gamma = 0$, indicating a \textit{structural} left-right asymmetry independent of non-Hermiticity. The origin of this offset lies in the bipartite hopping structure: intracell hopping ($\alpha \leftrightarrow \beta$ within a unit cell) differs from intercell hopping ($\beta_j \to \alpha_{j+1}$), so observation sites on different sublattices experience inequivalent coupling pathways. Despite these structural offsets, all curves show $d\Delta_{\mathbb{T}}/d\gamma > 0$ for small values of $|\gamma|$: increasing $\gamma$ always enhances $\mathbb{T}_{R \to L}$ relative to $\mathbb{T}_{L \to R}$---a universal trend reflecting the underlying non-reciprocal hopping. To isolate the genuine non-Hermitian scissors effect from structural artifacts, all subsequent results use the $\alpha\alpha\alpha$ configuration.

\paragraph{Three temporal regimes.}
\label{sec:phases}
The time evolution exhibits three distinct regimes (Fig.~\ref{fig:scissors}): (I) At early times ($t \lesssim 4$), QLIF remains negligible until information propogates to observation sites, governed by the Lieb-Robinson bound. For $\gamma \neq 0$, the NHSE creates asymmetric onset times---the earliest dynamical signature of non-reciprocity. (II) After the onset transient ($4 \lesssim t \lesssim 10$), QLIF enters a quasi-stationary regime where $\gamma$-dependent asymmetry is most clearly resolved. A sign rule emerges: $\text{sgn}(I_\gamma) \approx -\text{sgn}(\gamma)$ for $|\gamma| \geq 0.3$, where $I_\gamma$ is the time-integrated QLIF, reflecting sublattice-level probability accumulation. (III) At late times ($t \gtrsim 10$), persistent oscillations appear with period $T_\text{osc} \approx 2\pi/\Delta E$ determined by intrinsic energy-level spacing, independent of system size (see Supplemental Material~\cite{SM}). The NHSE protects oscillation amplitude against dilution in large systems because skin-localized eigenstates maintain high local probability density.

\begin{figure}[htbp]
    \centering
    \includegraphics[width=0.95\columnwidth]{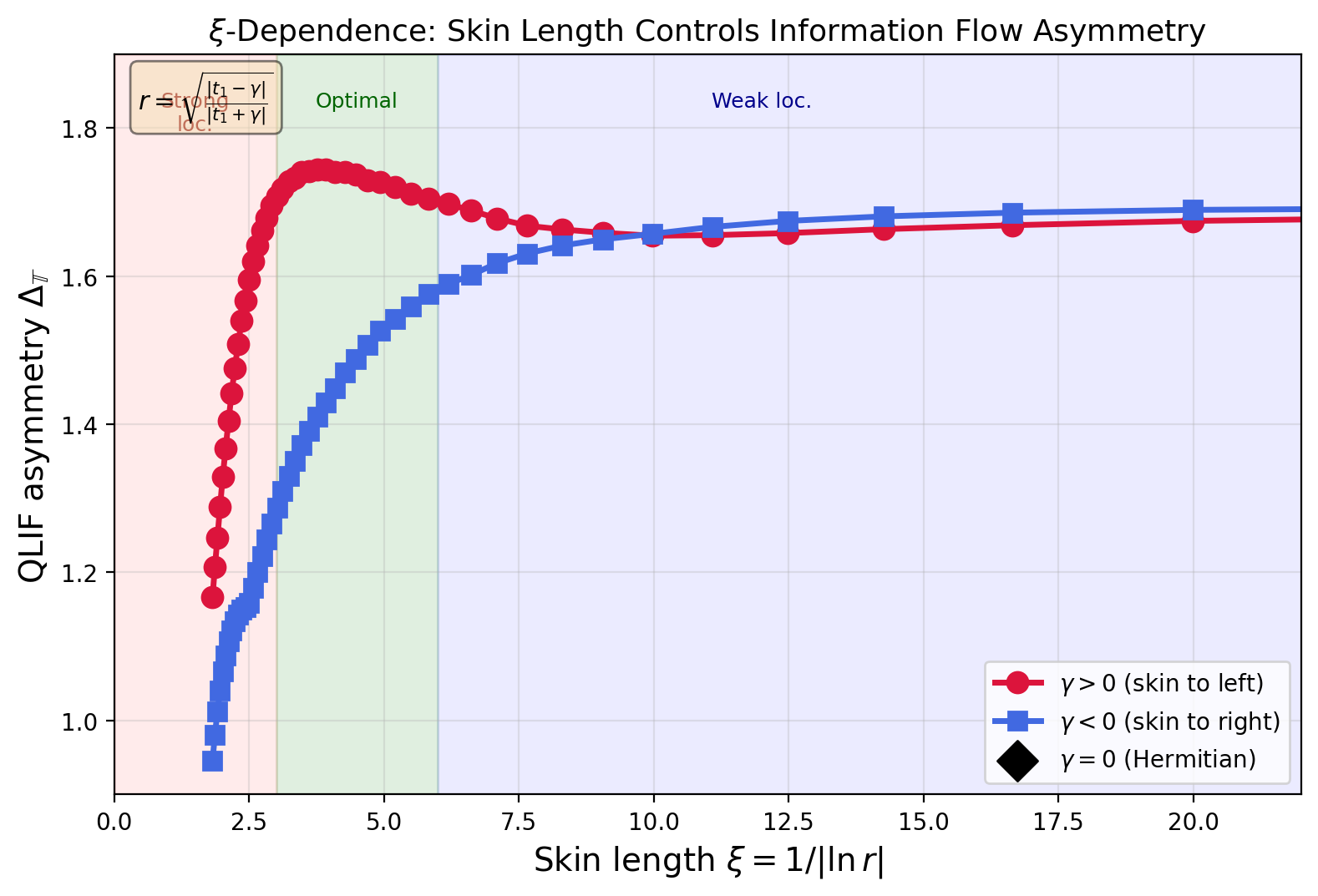}
    \caption{QLIF asymmetry $\Delta_{\mathbb{T}}$ versus skin length $\xi$ for $\gamma > 0$ (red circles) and $\gamma < 0$ (blue squares). Black diamond: Hermitian limit. The non-monotonic dependence reflects competition between symmetry breaking (favoring small $\xi$) and signal suppression (when eigenstates are too boundary-localized). Three regimes are visible: strong localization (left), optimal asymmetry (center), and weak localization approaching the Hermitian limit (right).}
    \label{fig:xi_unified}
\end{figure}

\paragraph{Skin length dependence.}
\label{sec:nhse_qlif}
While $\gamma$ directly enters the Hamiltonian, the physically relevant quantity controlling eigenstate localization is the skin length $\xi = 1/|\ln r|$, where $r = \sqrt{|(t_1 - \gamma)/(t_1 + \gamma)|}$ determines the exponential envelope $|\psi_n| \sim r^n$ of skin-localized eigenstates. Figure~\ref{fig:xi_unified} reveals a central result: the QLIF asymmetry $|\Delta_{\mathbb{T}}|$ exhibits \textit{non-monotonic} dependence on $\xi$, reaching a maximum at $\xi_\text{opt} \approx 3$--$4$ before declining.

The physical mechanisms can be understood through three regimes: In the \textit{strong localization regime} (small $\xi$), eigenstates are excessively confined near boundaries, suppressing signal propagation to bulk observation sites---paradoxically, extreme NHSE confines dynamics to boundaries and reduces the measurable effect. In the \textit{optimal regime} (moderate $\xi$ comparable to observation distance), non-reciprocity-induced symmetry breaking is strong while bulk transport remains viable, producing maximal directional asymmetry. In the \textit{weak localization regime} (large $\xi$), the system approches the Hermitian limit and directional signatures diminish. The $\gamma < 0$ and $\gamma > 0$ branches show different peak positions, reflecting the interplay between skin direction and measurement geometry. This non-monotonic behavior establishes the first quantitative link between static skin localization length and dynamic information transfer, implying that \textit{moderate} non-Hermiticity produces the strongest QLIF signatures.

Figure~\ref{fig:xi_unified} displays $\Delta_{\mathbb{T}}$ as a function of $\xi$ for both signs of $\gamma$. The $\gamma > 0$ branch (red circles, skin localization toward left boundary) and $\gamma < 0$ branch (blue squares, skin localization toward right boundary) both show non-monotonic behavior but with different peak positions and amplitudes. This asymmetry between branches reflects the interplay between skin direction and the measurement geometry: when skin localization aligns with the direction being probed, the signal is enhanced; when opposed, it is suppressed. The black diamond marks the Hermitian limit ($\gamma = 0$, $\xi \to \infty$), where $\Delta_{\mathbb{T}} = 0$ by symmetry. The three shaded regions indicate the physical regimes discussed above.

\paragraph{NHSE blocking and velocity ordering.}
\label{sec:lightcone}
To probe the causal structure of information propagation, we measure the onset time $t^*$---the earliest time at which $|\mathbb{T}_{j_0 \to j_0+d}| > \epsilon$ ($\epsilon = 10^{-6}$)---as a function of distance $d$. The frozen site is placed at the initial excitation $j_0 = L/2$, so that QLIF directly measures causal influence from the source. Two theoretical bounds constrain propagation: the maximum group velocity $v_\text{max} = 2\min(t_1, t_2) = 1.0$ and the Lieb-Robinson bound $v_\text{LR} = 2\max(t_1, t_2) = 2.0$.

\begin{figure}[htbp]
    \centering
    \includegraphics[width=0.95\columnwidth]{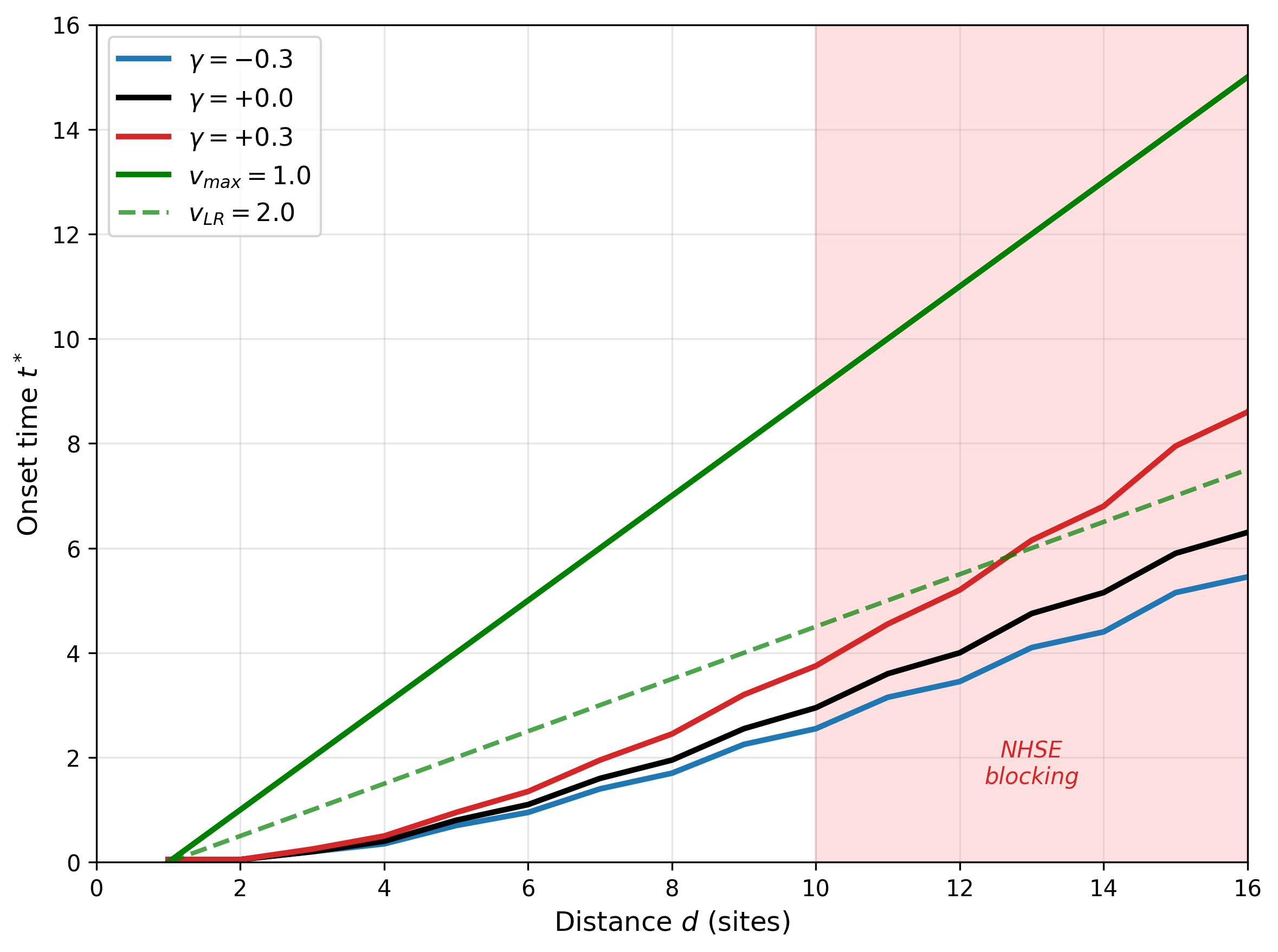}
    \caption{Light-cone analysis. Onset time $t^*$ versus distance $d$ for $\gamma = -0.3$ (blue), $0$ (black), $+0.3$ (red). Solid green: $v_\text{max} = 1.0$; dashed green: $v_\text{LR} = 2.0$. The shaded region ($d > 10$) marks NHSE-induced blocking where $\gamma > 0$ strongly deviates from linear scaling. Parameters: $L = 42$, $t_1 = 1$, $t_2 = 0.5$.}
    \label{fig:lightcone}
\end{figure}

A clear $\gamma$-dependent velocity ordering emerges (Fig.~\ref{fig:lightcone}):
\begin{equation}
    v_{\mathrm{eff}}(\gamma < 0) > v_{\mathrm{eff}}(0) > v_{\mathrm{eff}}(\gamma > 0).
    \label{eq:velocity_order}
\end{equation}
This hierarchy directly reflects the NHSE blocking phenomenon: $\gamma > 0$ localizes eigenstates toward the left boundary, impeding rightward propagation. Modes travelling \textit{against} the skin direction experience exponential suppression $\sim e^{-d/\xi}$ over the localization length, while those aligned with the skin direction propagate more freely. In the non-blocking regime ($d \lesssim 10$), data points lie close to the Lieb-Robinson bound $v_\text{LR} = 2.0$, with linear fits yielding effective velocities $v_{\mathrm{eff}} \approx 1.8$--$2$. At large distances ($d \gtrsim 10$, shaded region), blocking becomes dramatic---$\gamma > 0$ shows a sharp upturn as information must propagate against the skin direction. This NHSE blocking provides a dynamical manifestation of the static eigenstate localization, demostrating that the skin effect not only reshapes the eigenstate distribution but fundamentally alters information transport.

Figure~\ref{fig:lightcone} displays the onset time $t^*$ as a function of distance $d$ for three representative $\gamma$ values. The two theoretical velocity bounds are shown: the maximum group velocity $v_\text{max} = 1.0$ (solid green) derived from the dispersion relation, and the Lieb-Robinson bound $v_\text{LR} = 2.0$ (dashed green). In the non-blocking regime ($d \lesssim 10$), all three curves lie between these bounds, with the $\gamma = 0$ case (black) serving as the symmetric reference. The velocity ordering is clearly visible: $\gamma = -0.3$ (blue) propagates fastest, followed by $\gamma = 0$, with $\gamma = +0.3$ (red) being slowest. The shaded region ($d > 10$) reveals the dramatic onset of NHSE blocking: while $\gamma \leq 0$ maintains approximately linear scaling, the $\gamma = +0.3$ curve shows a sharp upturn, indicating that information propagation against the skin direction becomes increasingly suppressed at large distances. This ``superluminal'' appearance at small $d$ (points slightly below $v_\text{LR}$) arises from our choice of a small threshold ($\epsilon = 10^{-6}$), which detects the exponentially suppressed signal that leaks beyond the causal light cone~\cite{LiebRobinson1972,Bravyi2006}.

\section{Discussion}
\label{sec:discussion}

Our results reveal how each observable connects to the underlying physics. The scissors effect (Fig.~\ref{fig:scissors}) demonstrates that QLIF directly captures the directional bias induced by non-reciprocal hopping. The full-range $\gamma$ scan (Fig.~\ref{fig:delta_gamma}) reveals non-monotonic behavior with a striking prediction: the strongest NHSE (largest $|\gamma|$) produces the \textit{weakest} measurable asymmetry. This occurs because as $|\gamma| \to t_1$, intracell hopping becomes completely unidirectional ($t_1 \pm \gamma \to 0$ or $2t_1$), confining all eigenstates to within 1--2 sites of the boundary. Consequently, excitations initialized in the bulk cannot propagate, and both $\mathbb{T}_{R \to L}$ and $\mathbb{T}_{L \to R}$ vanish. The optimal asymmetry at moderate $|\gamma| \approx 0.15$--$0.3$ reflects a balance: non-reciprocity is strong enough to break symmetry, yet skin localization remains weak enough to permit bulk transport. This non-monotonic behavior, consistently observed in both the $\gamma$-scan (Fig.~\ref{fig:delta_gamma}) and $\xi$-scan (Fig.~\ref{fig:xi_unified}), provides clear experimental guidance: moderate non-Hermiticity maximizes observable signatures. The sublattice analysis (Fig.~\ref{fig:sublattice}) distinguishes genuine non-Hermitian effects from structural artifacts. The velocity ordering (Fig.~\ref{fig:lightcone}) provides direct evidence of NHSE blocking.

QLIF overcomes limitations of conventional measures: unlike symmetric correlation functions, it captures the asymmetry $\mathbb{T}_{R \to L} \neq \mathbb{T}_{L \to R}$ characteristic of non-reciprocal dynamics. While Lieb-Robinson bounds~\cite{LiebRobinson1972,Bravyi2006} rely on quasiparticle propagation~\cite{Calabrese2005}---problematic for non-Hermitian systems with biorthogonal eigenstates~\cite{Kunst2018}---QLIF operates directly at the level of entropy production~\cite{ChenZhouChenYe2024CPL}. Our results extend recent Hermitian applications~\cite{Ghosh2025} to the non-Hermitian regime.

We note that our analysis relies on the standard correspondence between the non-reciprocity parameter $\gamma$ and the skin accumulation direction in the one-dimensional SSH model~\cite{YaoWang2018}. However, in more general settings---such as higher-dimensional lattices, photon-mediated interactions, or waveguide-coupled arrays---the direction of eigenstate accumulation does not always follow that of the asymmetric couplings, and skin effect \textit{reversal} can occur~\cite{YangLee2025,Roccati2024,ShiPoddubny2025}. Investigating how QLIF signatures---particularly the sign rule Eq.~(\ref{eq:sign_rule_delta}) and the velocity ordering Eq.~(\ref{eq:velocity_order})---are modified in systems exhibiting such reversed NHSE constitutes an intriguing direction for future work.

Several other directions remain open: interactions may modify the competition between NHSE localization and many-body delocalization; the interplay with nontrivial topology near $t_2 = t_1$~\cite{Li2020} merits investigation. Experimental verification appears feasible in photonic lattices~\cite{Weidemann2020,Wang2021Science} and topolectrical circuits~\cite{Hofmann2020,Helbig2020}. Extensions to many-body systems and higher-order skin effects~\cite{Zeng2022} would further illuminate directional dynamics.


\section{Conclusion}
\label{sec:conclusion}

We have applied quantum Liang information flow to the non-Hermitian SSH model, establishing the first quantitative connection between skin localization and directional information dynamics. Our principal findings are: (i) a scissors effect with $\Delta_{\mathbb{T}} \propto \gamma$ for small $|\gamma|$; (ii) non-monotonic $\xi$-dependence with optimal asymmetry at moderate localization; (iii) velocity ordering $v_{\mathrm{eff}}(\gamma < 0) > v_{\mathrm{eff}}(0) > v_{\mathrm{eff}}(\gamma > 0)$ demonstrating NHSE blocking; and (iv) three distinct temporal regimes. These results establish QLIF as a powerful probe for non-Hermitian dynamics, opening new avenues for understanding directional information propagation in non-reciprocal quantum systems.

\ack
Y.~B. acknowledges support from National Natural Science Foundation of China (Grant No.~12404551) and the China Postdoctoral Science Foundation (Grant No.~2024M750339).

\section*{References}
\bibliographystyle{unsrt}
\bibliography{references}

\end{document}